\documentclass{JHEP3}
\usepackage{amsmath}
\usepackage{amssymb}
\newcommand{\cN}{\mathcal{N}}

\def\HBS{\mathbb S}
\def\HS{S}
\def\reciP{\mathcal{P}}

\newcommand{\beq}{\begin{equation}}
\newcommand{\eeq}{\end{equation}}
\newcommand{\beqa}{\begin{eqnarray}}
\newcommand{\eeqa}{\end{eqnarray}}

\newcommand{\M}{M}

\def\ztw{\zeta_2}
\def\zt{\zeta_3}
\def\zfr{\zeta_4}
\def\zf{\zeta_5}
\def\zsx{\zeta_6}
\def\zs{\zeta_7}
\def\ze{\zeta_8}
\def\zn{\zeta_9}
\title{Double-logs, Gribov-Lipatov reciprocity and wrapping}

\author{V.~N.~Velizhanin\\ %}
Theoretical Physics Department\\
Petersburg Nuclear Physics Institute\\
Orlova Roscha, Gatchina\\
188300 St.~Petersburg, Russia\\ %}
E-mail: \email{velizh@thd.pnpi.spb.ru}}
\abstract{
We study analytical properties of the five-loop anomalous dimension of twist-2 operators at negative even values of Lorentz spin. Following L.~N.~Lipatov and A.~I.~Onishchenko, we have found two possible generalizations of double-logarithmic equation, which allow to predict a lot of poles of anomalous dimension of twist-2 operators at all orders of perturbative theory from the known results. Second generalization is related with the reciprocity-respecting function, which is a single-logarithmic function in this case. We have found, that the knowledge of first orders of the  reciprocity-respecting function gives all-loop predictions for the highest poles. Obtained predictions can be used for the reconstruction of a general form of the wrapping corrections for twist-2 operators.
}

\begin{document}

\maketitle

\section{Introduction}
\label{sec:intro}

The anomalous dimension of composite gauge-invariant operators in $\cN=4$ SYM theory can be calculated with the help of integrability. Integrability in the context of AdS/CFT-correspondence~\cite{Maldacena:1997re,Gubser:1998bc,Witten:1998qj} was found from the investigation of BMN-operators~\cite{Berenstein:2002jq} in the leading order of perturbative theory in ref.~\cite{Minahan:2002ve}.\footnote{Earlier, the similar integrability was opened in quantum chromodynamics in the Regge limit~\cite{Lipatov:1993yb,Lipatov:1994xy,Faddeev:1994zg} and for some of operators~\cite{Braun:1998id}.}
Generalization to the higher orders together with the investigations of the integrable structures from the superstring theory side, started in ref.~\cite{Bena:2003wd}, allowed to formulate all-loop asymptotic Bethe equations~\cite{Beisert:2003tq}--\cite{Beisert:2006ez}.
%equations~\cite{Beisert:2003tq,Beisert:2003yb,Beisert:2003jj,Beisert:2003ys,Serban:2004jf,Kazakov:2004qf,Beisert:2004hm,Arutyunov:2004vx,Staudacher:2004tk,Beisert:2005fw,Beisert:2005tm,Janik:2006dc,Hernandez:2006tk,Arutyunov:2006iu,Beisert:2006ib,Eden:2006rx,Bern:2006ew,Beisert:2006ez}.
For the operators with finite length the asymptotic Bethe equations will give non-complete result due to appearance of wrapping effects~\cite{SchaferNameki:2006ey,Kotikov:2007cy}.
The computations of wrapping corrections can be performed with integrability again~\cite{Bajnok:2008bm}--\cite{Bajnok:2010ud}.
%again~\cite{Bajnok:2008bm,Bajnok:2008qj,Bajnok:2009vm,Lukowski:2009ce,Gromov:2009tv,Gromov:2009bc,Gromov:2009zb,Bombardelli:2009ns,Arutyunov:2009ax,Arutyunov:2009ur,Arutyunov:2010gb,Balog:2010xa,Bajnok:2010ud}.
Independent tests of the obtained results were performed with the perturbative calculations in refs.~\cite{Fiamberti:2007rj,Fiamberti:2008sh,Velizhanin:2008jd,Velizhanin:2008pc}.
Moreover, there are all-loop predictions coming from the Balitsky-Fadin-Kuraev-Lipatov (BFKL)~\cite{Lipatov:1976zz,Kuraev:1977fs,Balitsky:1978ic}
and
double-logarithmic~\cite{Gorshkov:1966ht}--\cite{Kirschner:1983di}
%double-logarithmic~\cite{Gorshkov:1966ht,Gorshkov:1966hu,Gorshkov:1966qd,Kirschner:1982qf,Kirschner:1982xw,Kirschner:1983di}
equations: analytical continuation of the anomalous dimension of twist-2 operators should matched with corresponding all-loop results.
However, we can do the inverse: we can use BFKL and double-logarithmic constraints for the reconstruction of wrapping corrections if the part coming from the asymptotic Bethe equations is already known.
We have tried such method at four-loop order~\cite{Kotikov:2007cy}, but at that time we did not have enough constraints and did not know a general structure of wrapping corrections for the twist-2 operators.
Moreover, we did not use such remarkable property of anomalous dimension as generalized Gribov-Lipatov reciprocity~\cite{Dokshitzer:2005bf,Dokshitzer:2006nm}.

The aim of this paper is analyze the general properties of the anomalous dimension of twist-2 operators. Investigations in this direction give us two new results, which provide a lot of constraints for the anomalous dimension of twist-2 operators. Both of these results are related with the generalized double-logarithmic equation.
For the first time, a similar equation was suggested by Lev N. Lipatov and Andrei Onishchenko at 2004, but was not published. Then, it was improved by Lev N. Lipatov in ref.~\cite{Kotikov:2007cy}.
In section~\ref{sec:DL} we explain the method of analytical continuation, which we used, together with the obtained result. The generalized double-logarithmic equation will be written at section~\ref{sec:GenerDL} and we give a remarkable consequence of this equation. Investigation of analytical continuation of the reciprocity-respecting function gives us one more remarkable result, which will be described in details  in section~\ref{sec:RRF}. In section~\ref{sec:Wrapping} we show, how obtained constraints can be used for the reconstruction of the wrapping corrections. In the end, we will discuss obtained results and perspective of their usage.

\section{Double-logarithmic equation} \label{sec:DL}

The double-logarithmic asymptotics of the scattering amplitudes were investigated in QED and QCD in the papers~\cite{Gorshkov:1966ht,Gorshkov:1966hu,Gorshkov:1966qd} and~\cite{Kirschner:1982qf,Kirschner:1982xw,Kirschner:1983di}
(see also {\texttt{arXiv}} version of ref.~\cite{Kotikov:2002ab}). It corresponds to summing the leading terms $(\alpha \ln^2s)^k$ in all orders of perturbation theory.
In combination with a Mellin transformation, the double-logarithmic asymptotics allow to predict the singular part of anomalous dimensions near the point $\M=-2$.
For our purpose and in our notation double-logarithmic equation has the following form
\begin{equation}\label{DL}
\gamma\,(2\,\omega+\gamma)=-16\,g^2\,.
\end{equation}
The solution of this equation gives prediction for the highest pole $(g^{2k}/\omega^{2k-1})$ in all orders of perturbative theory:
\begin{equation}
\gamma=-\omega+\omega\, \sqrt{1-\frac{16 g^2}{\omega^2}}
=
2\,\frac{(-4\, g^2)}{\omega}
-2\,\frac{(-4\, g^2)^2}{\omega^3}
+4\,\frac{(-4\, g^2)^3}{\omega^5}
-10\,\frac{(-4\, g^2)^4}{\omega^7}
%+28\,\frac{(-4\, g^2)^5}{\omega^9}
+\ldots\, .\label{DLSolve}
\end{equation}

To study double-logarithmic equation~(\ref{DL}) at higher-loop orders we need an analytical continuation of known result for the anomalous dimension of twist-2 operators.
Analytical continuation can be performed in the following way. First of all we translate the five-loop anomalous dimension of twist-2 operators~\cite{Kotikov:2003fb,Kotikov:2004er,Kotikov:2007cy,Bajnok:2008qj,Lukowski:2009ce}\footnote{Three-loop anomalous dimension was obtained from the result of perturbative calculation in QCD~\cite{Moch:2004pa} with the maximal transcendentality principle~\cite{Kotikov:2002ab}.} into canonical basis with the help of {\texttt{SUMMER}} package~\cite{Vermaseren:1998uu} for {\texttt{FORM}}~\cite{Vermaseren:2000nd}. Then, we use {\texttt{FORM}} package {\texttt{HARMPOL}}~\cite{Remiddi:1999ew} for the inverse Mellin transformation of obtained expressions and for extraction of $(\ln x)^k$ terms. At the end with the help of {\texttt{MATHEMATICA}} package {\texttt{HPL}}~\cite{Maitre:2005uu} we make the Mellin transformation into negative value $M=-2+\omega$.
The result can be written as
\begin{equation}\label{gammaDL}
\gamma_{\mathrm{DL}}(\omega)=\sum_{k=1}{\mathbb{D}}^k(\omega)\, g^{2k}=\sum_{k=1}\sum_{m=0}{\mathbb{D}}^k_m\,\omega^{-2k+1+m}\, g^{2k}\,,
\end{equation}
where we keep first eight terms in the $\omega$-expansion of ${\mathbb{D}}^k(\omega)=\sum_{m=0}{\mathbb{D}}^k_{m}\,\omega^{-2k+1+m}$, which can be found in eq.~\eqref{App:1} of Appendix.

\section{Generalized double-logarithmic equation} \label{sec:GenerDL}

Let us analyze obtained expression.
First of all we have substituted the result for the analytically continued anomalous dimension~(\ref{gammaDL}) into double-logarithmic equation~(\ref{DL}) and we have found the following general expression for the generalized double-logarithmic equation at $M=-2+\omega$ (we have omitted higher terms in $\omega$ for all powers of $g$):\footnote{For the first time, a similar equation was suggested by Lev N. Lipatov and Andrei Onishchenko at 2004,
but was not published. Then, it was improved by Lev N. Lipatov in ref.~\cite{Kotikov:2007cy}.}
\begin{eqnarray}
\gamma\,(2\,\omega+\gamma)&=&
-16\,g^2
-64\, g^4\, \ztw
+128\,g^6 \Big(\zt+2\, \zfr\Big)
+32\,g^8\! \bigg(80\,\ztw \zt+12\,\zt^2-4\,\zf+\frac{59}{3}\zsx\bigg)\nonumber\\&&
+256\,g^{10} \!
\bigg(
76\,\zt\, \zfr
-148\,\ztw\,\zf
-45\, \zs
\bigg)+\ldots\,,\label{DLgenerExp}
\end{eqnarray}
where $\zeta_i$ is Riemann zeta function and we keep terms up to transcendentality level $7$.
Surprisingly,
the
right hand side of last equation does not contain poles in $\omega$.
So, we suggest, that the generalized double-logarithmic equation at $M=-2+\omega$ have the following form:
\begin{equation}\label{DLgener}
\gamma\,(2\,\omega+\gamma)=\sum_{k=1}\sum_{m=0}{\mathfrak C}_m^k\,\omega^m\,g^{2k}\,,
\end{equation}
where coefficients are listed in eq.~\eqref{App:2} of Appendix up to five-loop order.

This result has very remarkable consequence.
If we solve last equation, we have found
\begin{equation}\label{DLgenerSolve}
\gamma_{\mathrm{DL}}(\omega)=-\omega+\sqrt{\smash[b]{\omega^2+\sum_{k=1}\sum_{m=0}{\mathfrak C}_m^k\,\omega^m\,g^{2k}}}\,.
\end{equation}
\vspace*{-3mm}

\noindent
Perturbatively expanding this solution we can predict in all orders of perturbative theory all poles up to $(g^2/\omega^2)^k\omega^{2\ell}$, if we know $\ell$-loop anomalous dimension (or first $\ell$ orders in right hand side of eq.~(\ref{DLgener})).

We can find also the relations between coefficients at poles if we substitute eq.~(\ref{gammaDL}) into eq.~(\ref{DLgener}):
\begin{eqnarray}
\sum_{k'=1}\ \sum_{m'=0}{\mathbb{D}}_{-2k'+1+m'}^{k'}\,\omega^{-2k'+1+m'}g^{2k'}\,
\Bigg(
2\,\omega &+&
\sum_{k''=1}\ \sum_{m''=0}{\mathbb{D}}_{-2k''+1+m''}^{k''}\,\omega^{-2k''+1+m''}g^{2k''}
\Bigg)=\nonumber\\[2mm]
&&\qquad\qquad=
\sum_{k=1}\sum_{m=0}{\mathfrak C}_m^k\,\omega^m\,g^{2k}\,.\label{DLgenerExpand}
\end{eqnarray}
Taken coefficients with the same powers of $g$ and $\omega$ we obtain:
\begin{equation}\label{DLgenerCoef}
{\mathbb{D}}_{-2k+1+m}^{k}=-\frac{1}{2}
\sum_{k'=1}^{k-1}\ \sum_{m'=0}^{m}
{\mathbb{D}}_{-2k'+1+m'}^{k'}\,
{\mathbb{D}}_{-2(k-k')+1+m-m'}^{k-k'}
+\frac{1}{2}{\mathfrak C}_{-2k+2+m}^k\,,
\end{equation}
where ${\mathfrak C}_{m}^k=0$ if $m<0$.

\section{Reciprocity respecting function} \label{sec:RRF}

Other interesting property comes from the consideration of the reciprocity-respecting function $\mathcal P(x)$~\cite{Dokshitzer:2005bf,Dokshitzer:2006nm}, which satisfies Gribov-Lipatov relation~\cite{Gribov:1972ri,Gribov:1972rt}:
\begin{equation}\label{GrLipRel}
{\mathcal P}(x)=-\,x\,{\mathcal P}\!\left(\frac{1}{x}\right)\,.
\end{equation}
In Mellin space the reciprocity-respecting function $\mathcal P(M)$ is related with the anomalous dimension through~\cite{Dokshitzer:2005bf,Dokshitzer:2006nm}
\begin{equation} \label{Pfunction5ll}
\gamma(M) = \reciP \left(M+\frac{1}{2} \gamma(M) \right)
\end{equation}
and can be found from the anomalous dimension with the help of the following equation~\cite{Basso:2006nk}
\begin{equation}\label{PfromGamma}
{\mathcal P}(M)=\sum_{k=1}^{\infty}\frac{1}{k!}\left(-\frac{1}{2}\,\frac{d}{dM}\right)^{k-1}\!\!\Big[\gamma(M)\Big]^k\,.
\end{equation}

The combination of the harmonic sums entering into expressions for the reciprocity respecting function ${\mathcal P}(M)$ for the twist-2 operators should satisfy some properties, which were proven in general in ref.~\cite{Beccaria:2009vt}.\footnote{From the first time the reciprocity was used for the reconstruction of the anomalous dimension for the twist-3 operators  at five-loop order~\cite{Beccaria:2009eq}.}
However, in our calculations~\cite{Lukowski:2009ce}, we did not use the reciprocity respecting sums, but instead we used the binomial harmonic sums, which can be expressed through the combinations of the usual harmonic sums (see~\cite{Lukowski:2009ce} for details).

\subsection{Analytical continuation at $M=-2+\omega$}\label{sec:RRF2}

Thus, in order to find the analytical continuation of the reciprocity respecting function at $M=-2+\omega$ we can use corresponding results for the anomalous dimension~\eqref{gammaDL} replacing $M$ with $\omega$ in eq.~\eqref{PfromGamma}. We have found, that the leading singularities of ${\mathcal P}_{\mathrm{DL}}(\omega)$ look like:
\begin{equation}\label{PDL}
{\mathcal P}_{\mathrm{DL}}(\omega)=\sum_{k=1}{\mathfrak{D}}^k\!\left(\frac{ g^{2}}{\omega}\right)^k\,,
\end{equation}
with
\begin{eqnarray}
{\mathfrak{D}}^1 &=& -8\,, \label{PDL1}\\
{\mathfrak{D}}^2 &=& 32\,, \label{PDL2}\\
{\mathfrak{D}}^3 &=& 128\,\ztw\,, \label{PDL3}\\
{\mathfrak{D}}^4 &=& -512 - 1536\, \ztw - 512\, \zt\,, \label{PDL4}\\
{\mathfrak{D}}^5 &=& 4096\, \ztw + 8192\, \zt - 8192\, \zfr\,.\label{PDL5}
\end{eqnarray}
This property can be proved in general, but we show, how it can be done in leading nontrivial case. We substitute eq.~(\ref{gammaDL}) into eq.~(\ref{PfromGamma}), expand it up to second order of perturbative theory and will interested only with the highest pole
\begin{eqnarray}
{\mathcal P}_{\mathrm{DL}}(\omega)&=&
\frac{1}{1!}\left({\mathbb{D}}^1_{-1}\frac{g^2}{\omega}\right)+
\frac{1}{2!}\left(-\frac{1}{2}\,\frac{d}{d\omega}\right)^{1}\left({\mathbb{D}}^1_{-1}\frac{g^2}{\omega}\right)^2+
\frac{1}{1!}\left({\mathbb{D}}^2_{-3}\frac{g^4}{\omega^3}\right)+\ldots\nonumber\\
&=&{\mathbb{D}}^1_{-1}\,\frac{g^2}{\omega}
-{\mathbb{D}}^1_{-1}{\mathbb{D}}^1_{-1}\,\frac{g^4}{4}\,\frac{d}{d\omega}\left(\frac{1}{\omega^2}\right)+
{\mathbb{D}}^2_{-3}\,\frac{g^4}{\omega^3}+\ldots\,.\label{highpolerem}
\end{eqnarray}
Using eq.~(\ref{DLgenerCoef}), which in our case gives ${\mathbb D}^2_{-3}=-{\mathbb D}^1_{-1}{\mathbb D}^1_{-1}/2$, one can verify, that the highest pole $g^4/\omega^3$ in eq.~(\ref{highpolerem}) is canceled.

The result~(\ref{PDL}) can be compared with the BFKL poles, i.e. with the anomalous dimension of twist-2 operators, which is analytically continued to $M=-1+\omega$:
\begin{equation}\label{BFKLLO}
\gamma_{\mathrm{BFKL}}(\omega)=\sum_{k=1}{\mathfrak{B}}^k\!\left(\frac{ g^{2}}{\omega}\right)^k\,,
\end{equation}
with
\begin{eqnarray}
{\mathfrak{B}}^1 &=& -8\,, \\
{\mathfrak{B}}^2 &=& 0\,, \\
{\mathfrak{B}}^3 &=& 0\,, \\
{\mathfrak{B}}^4 &=& -1024 \zt\,, \\
{\mathfrak{B}}^5 &=& 0\,.
\end{eqnarray}
In both case~(\ref{PDL}) and (\ref{BFKLLO}) we have the same single-logarithmic behavior of leading poles.

Such simple structure of reciprocity-respecting function ${\mathcal P}_{\mathrm{DL}}(\omega)$ in eq.~(\ref{PDL}) gives one more unexpected result: at any orders of perturbative theory the leading poles in $\omega$ near $M=-2+\omega$ is defined through one-loop ${\mathcal P}_{\mathrm{DL}}(\omega)$! Indeed, substitute one-loop reciprocity-respecting function from eqs.~(\ref{PDL}) and~(\ref{PDL1}) into equation ({cf.} with eq.~(\ref{PfromGamma}))
\begin{equation}\label{GammafromP}
\gamma(\omega)=\sum_{k=1}^{\infty}\frac{1}{k!}\left(\frac{1}{2}\,\frac{d}{d\omega}\right)^{k-1}\!\!\Big[{\mathcal P}(\omega)\Big]^k
\end{equation}
 we obtain
\begin{eqnarray}
\gamma_{\mathrm{DL}}^{\mathrm{LA}}(\omega)&=&\sum_{k=1}^{\infty}\frac{1}{k!}\left(\frac{1}{2}\,\frac{d}{d\omega}\right)^{k-1}\!\!\left[\frac{-8\,g^2}{\omega}\right]^k
=2\,\omega\sum_{k=1}^{\infty}\left(\frac{4\,g^2}{\omega^2}\right)^k\frac{\big(2\,(k-1)\big)!}{(k-1)!\,k!}\nonumber\\
&=&2\,\omega\sum_{k=1}^{\infty}\left(\frac{4\,g^2}{\omega^2}\right)^k\frac{\tbinom{2\,(k-1)}{k-1}}{k}
=-\omega+\omega\,\sqrt{1-\frac{16\,g^2}{\omega^2}}\,,\label{GammaDLfromPLO}
\end{eqnarray}
which is the solution~(\ref{DLSolve}) of the original double-logarithmic equation~(\ref{DL}).

For the next-to-leading poles we should take into account, that ${\mathfrak D}^k$ in eq.~(\ref{PDL}) is the function of $\omega$ and it is necessary keep the next term for ${\mathfrak D}^1$ in eq.~(\ref{PDL1}), which is equal to $8\,\omega$, while for ${\mathfrak D}^2$ from eq.~(\ref{PDL2}) first term is enough. For this case we can write eq.~(\ref{GammafromP}) as
\begin{equation}
\gamma_{\mathrm{DL}}^{\mathrm{NLA}}(\omega)=\sum_{k=1}^{\infty}\frac{1}{k!}\left(\frac{1}{2}\,\frac{d}{d\omega}\right)^{k-1}
\!\!\left[\frac{-8\,g^2(1-\omega)}{\omega}+\frac{32\,g^4}{\omega^2}\right]^k\,.
\end{equation}
Now, we extract from the above equation all terms, which is proportional to $\left({g^2}/{\omega^2}\right)^k\omega^2$ and simplify
\begin{eqnarray}
\gamma_{\mathrm{DL}}^{\mathrm{NLA}}(\omega)
&=&\sum_{k=1}^{\infty}\frac{1}{k!}\left(\frac{1}{2}\,\frac{d}{d\omega}\right)^{k-1}
\left[k\,\left(\frac{-8\,g^2}{\omega}\right)^{k-1}\left(\frac{32\,g^4}{\omega^2}\right)
+k\,\left(\frac{-8\,g^2}{\omega}\right)^{k-1}(8\,g^2\omega)\right]\nonumber\\
&=&2\,\omega^2\sum_{k=1}^{\infty}\left(\frac{4\,g^2}{\omega^2}\right)^k\frac{\big(2\,(k-1)\big)!}{(k-1)!\,(k-1)!}\nonumber\\
&=&2\,\omega^2\sum_{k=1}^{\infty}\left(\frac{4\,g^2}{\omega^2}\right)^k{\dbinom{2\,(k-1)}{k-1}}
=-\omega+\omega\,\sqrt{1-\frac{16\,g^2\,(1-\omega)}{\omega^2}}\,.\label{GammaDLfromP2}
\end{eqnarray}
The last equation is the solution of the original double-logarithmic equation~(\ref{DL}) with the right hand side multiplied by $(1-\omega)$, that is the generalized double-logarithmic equation~(\ref{DLgenerExp}).
We can go further and can obtain all loop results for $\left(g^2/\omega^2\right)^k\omega^{\ell}$ terms from the first $\ell$ orders of the reciprocity-respecting function ${\mathcal P}_{\mathrm{DL}}(\omega)$.

\subsection{Analytical continuation at $M=-r+\omega\,,\ r=2,4,6,\ldots$}\label{sec:RRFr}

We have found that eq.~(\ref{PDL}) hold true not only for $M=-2+\omega$, but for all other negative even values $M=-r+\omega\,,\ r=2,4,6,\ldots$\ .
In this case the anomalous dimension can be written as
\begin{equation}
\gamma_{\mathrm{DL}}(\omega,r) =2\,\sum _{k=1}{\mathbb{D}}^k(\omega,r)\,
\left(-4\,g^2\right)^k\,,\label{DLrgener}
\end{equation}
with (cf. v5 of the {\texttt {arXiv}} version of \cite{Kotikov:2004er} and~\cite{Kotikov:2007cy})
\begin{eqnarray}
{\mathbb{D}}^1(\omega,r)&=&\frac{1}{\omega}-\HS_1-\omega(\ztw+\HS_2)+\omega(\zt-\HS_3)+...\,,\nonumber\\
{\mathbb{D}}^2(\omega,r)&=&-\frac{1}{\omega^3}+\frac{2\,\HS_1}{\omega^2}+
\frac{\zeta_2+\HS_2}{\omega}-\frac{1}{2}\left(\zt-\HS_{-3}-2\HS_{-2}\HS_{1}-2\HS_{2}\HS_{1}+\HS_3+2\HS_{-2,1}\right)+...\,,\nonumber\\
{\mathbb{D}}^3(\omega,r)&=&\frac{2}{\omega^5}-\frac{6\,\HS_1}{\omega^4}+
\frac{-4\,(\ztw+\HS_2)+4\,\HS_1^2+(\HS_2+\HS_{-2})}{\omega ^3}\nonumber\\&&
+\frac{2 \HS_{-2,1}+2 S_1 \left(-2 \HS_{-2}+4 \HS_2+6 \ztw\right)+{3 \HS_{-3}}+{\HS_3}+6 \zt}{2\omega^2}
+...\,,\nonumber\\
{\mathbb{D}}^4(\omega,r)&=&-\frac{5}{\omega ^7}+\frac{20\,\HS_1}{\omega ^6}+
\frac{14\,(\zeta _2+\HS_2)-24\,\HS_1^2-4\,(\HS_2+\HS_{-2})}{\omega ^5}\nonumber\\&&
+\frac{8 \HS_1^3-4 \HS_{-2,1}-12 \HS_1 \left(2 \HS_2-\HS_{-2}+3 \ztw\right)-4 \HS_{-3}-11 \zt}{\omega^4}
+...\,,
\label{dlp}
\end{eqnarray}
where $\HS_i=\HS_i(r-1)$.
Using eq.~(\ref{PfromGamma}) we can translate this result for anomalous dimension into reciprocity-respecting function
\begin{equation}
{\mathcal P}_{\mathrm{DL}}(\omega,r) =2\,\sum _{k=1}{\mathcal{D}}^k(\omega,r)
\left(-4\,g^2\right)^k\,,
\end{equation}
with the following coefficients
\begin{eqnarray}
{\mathcal{D}}^1(\omega,r)&=&\frac{1}{\omega}-\HS_1-\omega(\ztw+\HS_2)+\omega^2(\zt-\HS_3)+...\,,\nonumber\\
{\mathcal{D}}^2(\omega,r)&=&-\frac{\HS_1}{\omega^2}-
\frac{\zeta_2+\HS_2}{\omega}-\frac{1}{2}\left(3\zt-\HS_{-3}-2\HS_{-2}\HS_{1}+2\ztw\HS_{1}-\HS_3+2\HS_{-2,1}\right)+...\,,\nonumber\\
{\mathcal{D}}^3(\omega,r)&=&\frac{\HS_1^2-\ztw+\HS_{-2}}{\omega ^3}
+\frac{S_1 \left(6 \ztw-2 \HS_{-2}+4 \HS_2\right)+4 \HS_{-3}+2 \HS_3+3 \zt}{2\,\omega^2}
+...\,,\nonumber\\
{\mathcal{D}}^4(\omega,r)&=&\frac{2\zt -2 \HS_1^3 + 2 \HS_{-2,1} - 6 \HS_1 \left(\ztw - \HS_{-2}\right)-\HS_{-3} - \HS_{3}}{2\,\omega^4}
+...\,.
\label{Plp}
\end{eqnarray}
We see again a single-logarithmic behavior of the reciprocity-respecting function ${\mathcal P}_{\mathrm{DL}}(\omega,r)$, i.e. it can be written as
\begin{equation}
{\mathcal P}_{\mathrm{DL}}(\omega,r) =2\,\sum _{k=1}\sum _{m=0}{\mathcal{D}}^k_m(r)\,\omega^m
\left(\frac{-4\,g^2}{\omega}\right)^k\,.\label{PDLr}
\end{equation}
We assume, that such form of reciprocity-respecting function ${\mathcal P}_{\mathrm{DL}}(\omega,r)$ near negative even values of $M$ will remain in all loops.

In our paper~\cite{Kotikov:2007cy} the following form of generalization of double-logarithmic equation was suggested:\footnote{Note again, that for the first time, a similar equation was suggested by Lev N. Lipatov and Andrei Onishchenko at 2004, but was not published. Then, it was improved by Lev N. Lipatov in ref.~\cite{Kotikov:2007cy}.}
\begin{eqnarray}
\label{dlnext}
\gamma\,(2\,\omega+\gamma)\ =\ &-&16\, g^2\left(1-\HS_1\,
\omega-(\HS_2+\zeta _2)\,\omega ^2\right)
-64\,g^4(\HS_2+\zeta _2-\HS_1^2)\nonumber\\
&-&4\,g^2\,
(\HS_2+\HS_{-2})\,\gamma^2\,,
\end{eqnarray}
which correctly reproduces the corresponding result in eq.~(\ref{DLrgener}). However, further generalization of equation~\eqref{dlnext} is not clear. Our equation~(\ref{PDLr}) gives regular method for obtaining a lot of predictions for poles of the anomalous dimension of twist-2 operators at negative~even~$M$. Note, that the generalized double-logarithmic equation~(\ref{dlnext}) for negative even values of~$M$ contains poles in the right hand side, which are written in the form of powers of $\gamma$. This fact violates our consideration of the generalized double-logarithmic equation~(\ref{DLgener}) at $M=-2+\omega$ and in the case of $M=-r+\omega$ with $r=4,6,8,\ldots$ we can predict all poles only up to $(g^2/\omega^2)^k\omega^{\ell}$, if we know $\ell$-loop result.

\section{Wrapping corrections reconstruction} \label{sec:Wrapping}

Let us suggest, that eqs.~(\ref{DLgener}) and~(\ref{PDLr}) will correct at all loops. This will give for us a lot of constraints to the anomalous dimension (or reciprocity-respecting function), which can be used for reconstruction of the wrapping corrections if ABA part is already known. Here, we show how it works at four loops. The result for ABA part can be found in ref.~\cite{Kotikov:2007cy}. The analytical continuation of this result into $M=-2+\omega$ has the following form
\begin{equation}\label{gammaDLABA}
\gamma^{\mathrm {ABA}}_{\mathrm{DL}}(\omega)=\sum_{k=1}{\overline{\mathbb{D}}}^k_m\,\omega^m\, g^{2k}\,,
\end{equation}
where coefficients $\overline{\mathbb{D}}^k_m$ can be found in eq.~\eqref{App:3} of Appendix.

There are $23$ predictions coming from the generalized double-logarithmic equation~(\ref{DLgener}), from equation~(\ref{PDLr}) for reciprocity-respecting function ${\mathcal P}_{\mathrm{DL}}(\omega)$ and from BFKL equation~(\ref{BFKLLO}). Using the maximal transcendentality principle~\cite{Kotikov:2002ab} and generalized Gribov-Lipatov reciprocity~\cite{Dokshitzer:2006nm} we can found, that at four loops we need $2^6+2^3+2^1=64+8+2=74$ binomial harmonic sums in the basis for the rational, $\zt$ and $\zf$ parts correspondingly, according to four-loop result for Konishi~\cite{Fiamberti:2007rj,Fiamberti:2008sh,Bajnok:2008bm,Velizhanin:2008jd}.
The binomial harmonic sums $\HBS_{i_1,\ldots,i_k}$ are defined through (see \cite{Vermaseren:1998uu})
\beq
\HBS_{i_1,\ldots,i_k}(M)=(-1)^M\sum_{j=1}^{M}(-1)^j\binom{M}{j}\binom{M+j}{j}\HS_{i_1,...,i_k}(j)\,,\qquad i_k>0\,,
\eeq
where $\HS_{i_1, \ldots ,i_k}$ are the nested harmonic sums (see \cite{Vermaseren:1998uu})
\beq \label{vhs}
\HS_a (M)=\sum^{M}_{j=1} \frac{(\mbox{sgn}(a))^{j}}{j^{\vert a\vert}}\, , \qquad
\HS_{a_1,\ldots,a_n}(M)=\sum^{M}_{j=1} \frac{(\mbox{sgn}(a_1))^{j}}{j^{\vert a_1\vert}}
\,\HS_{a_2,\ldots,a_n}(j)\, .
\eeq
However, if we suggest, that the wrapping effects can be imagine as a system of two spin chain, we will need consider the binomial sums, multiplied by $\HBS_1^2$ (see ref.~\cite{Bajnok:2008qj}), that lowers the level of transcendentality.\footnote{Such common factor was found also from the direct perturbative calculation of leading transcendental contribution to the four-loop anomalous dimension of twist-2 operator in ref.~\cite{Velizhanin:2008pc}.} There are $2^4+2^1+2^0=16+2+1=19$ binomail harmonic sums in the corresponding basis:
\begin{eqnarray}
\HBS^2_1\,&\!\!\Big\{\!\!&\HBS_5,\HBS_{4,1},\HBS_{3,2},\HBS_{2,3},\HBS_{1,4},\HBS_{3,1,1},\HBS_{2,2,1},\HBS_{2,1,2},\HBS_{1,3,1},\HBS_{1,2,2},\HBS_{1,1,3},\nonumber\\
&&\qquad\HBS_{2,1,1,1},
\HBS_{1,2,1,1},\HBS_{1,1,2,1},\HBS_{1,1,1,2},\HBS_{1,1,1,1,1},\HBS_{2}\,\zt,\HBS_{1,1}\,\zt,\zf\Big\}\label{BSBasisL4F}
\end{eqnarray}
and coefficients can be fixed uniquely.
Note, that with generalized double-logarithmic equation~(\ref{DLgener}) we can control even $\zf$ term, what is impossible with BFKL equation~(\ref{BFKLLO}).

\section{Conclusion} \label{sec:discussion}

In this paper using analytical continuation of the five-loop anomalous dimension of \mbox{twist-2} operators in $\cN=4$ SYM theory we have found some new remarkable properties of anomalous dimension.

Firstly, we have found, that the original double-logarithmic equation~(\ref{DL}) is modified in the right hand side only with regular over $\omega$ terms~\eqref{DLgener}.
This fact allows predict all-loop poles at $M=-2+\omega$ from the $\ell$-loop results up to $(g^2/\omega^2)^k\omega^{2\ell}$ terms~(\ref{DLgenerSolve}).

Secondly, we have found, that the reciprocity-respecting function ${\mathcal P}_{\mathrm{DL}}(\omega)$ for the analytically continued five-loop anomalous dimension of twist-2 operators at $M=-2+\omega$ behaves as single-logarithmic function~(\ref{PDL}), similar to the behavior of the anomalous dimension near BFKL poles.
This property allows to find leading poles at any loop orders from the first terms of ${\mathcal P}_{\mathrm{DL}}(\omega)$: the leading pole (original double-logarithmic pole) can be found from the leading order of ${\mathcal P}_{\mathrm{DL}}(\omega)$, the next-to-leading pole $(g^2/\omega^2)^k\omega^2$ can be found from the first two orders of ${\mathcal P}_{\mathrm{DL}}(\omega)$, the next-next-to-leading pole $(g^2/\omega^2)^k\omega^3$ can be found from the first three orders of ${\mathcal P}_{\mathrm{DL}}(\omega)$ and so on.
For $M=-2+\omega$ this property does not give any new information with compare to generalized double-logarithmic equation~(\ref{DLgenerCoef}), but we check, that the equation~(\ref{PDL}) will hold for all other negative even $M=-r+\omega,\ r=2,4,6,\ldots\ $, which gives a lot of new predictions for the pole structure of the anomalous dimension (or constraints for the reciprocity-respecting function) near negative even $M$.

Up to now we known a small part of analytical structure of the anomalous dimension of twist-2 operators from BFKL equation, which is calculated up to next-to-leading approximation~\cite{Fadin:1998py,Kotikov:2000pm}.
In this paper we have found corrections to the double-logarithmic equation from the analysis of the analytically continued five-loop anomalous dimension of twist-2 operators without calculations of any corrections to the double-logarithmic equation itself. It will be interesting to find a similar method for BFKL equation too.

In the future, we are going to try to reconstruct the six-loop anomalous dimension of twist-2 operators in $\cN=4$ SYM theory using generalized Gribov-Lipatov reciprocity and constraints from BFKL equation and generalized double-logs.
Note again, that with eq.~\eqref{DLgener} we can control all transcendental structures at six loops from the five-loop result. But first of all we should compute ABA part of anomalous dimension.

%%%%%%%%%%%%%%%%%%%%%%%%%%%%%%%%%%%%%%%%%%%%%%%%%%%%%%%%%%%%%%%%%%

\acknowledgments

I would like to thank to Andrei Onishchenko and Lev Nikolaevich Lipatov for the fruitful and stimulated discussions. This work is supported by RFBR grants 10-02-01338-a, RSGSS-65751.2010.2. % 3628.2008.2.

%%%%%%%%%%%%%%%%%%%%%%%%%%%%%%%%%%%%%%%%%%%%%%%%%%%%%%%%%%%%%%%%%%
\newpage
\appendix
%%%%%%%%%%%%%%%%%%%%%%%%%%%%%%%%%%%%%%%%%%%%%%%%%%%%%%%%%%%%%%%%%%
\section{Coefficients of analytical continuation of anomalous dimension} \label{SpecialSums}

In Appendix we are listed the coefficients in the equations~(\ref{gammaDL}), (\ref{DLgener}) and~(\ref{gammaDLABA}).

The coefficients in eq.~(\ref{gammaDL}) are equal to:
\begin{eqnarray}\label{App:1}
\mathbb{D}^1_{-1}&=&-8,\qquad
\mathbb{D}^1_{0}=8,\qquad
\mathbb{D}^1_{1}=8+8 \ztw,\qquad
\mathbb{D}^1_{2}=8-8 \zt,\qquad
\mathbb{D}^1_{3}=8+8 \zfr,\nonumber\\
\mathbb{D}^1_{4}&=&8-8 \zf,\qquad
\mathbb{D}^1_{5}=8+8 \zsx,\qquad
\mathbb{D}^1_{6}=8-8 \zs,\nonumber\\
\mathbb{D}^2_{-3}&=&-32,\qquad
\mathbb{D}^2_{-2}=64,\qquad
\mathbb{D}^2_{-1}=32 \ztw + 32,\qquad
\mathbb{D}^2_{0}=-16 \zt, \nonumber\\
\mathbb{D}^2_{1}&=&32 \ztw
-16 \zt
-20 \zfr
-32, \nonumber\\
\mathbb{D}^2_{2}&=&
-64 \zt \ztw
+64 \ztw
-48 \zt
+12 \zfr
+116 \zf
-64, \nonumber\\
\mathbb{D}^2_{3}&=&
40 \zt^2
-80 \zt
+96 \ztw
+44 \zfr
-8 \zf
-({115}/{3}) \zsx
-96, \nonumber\\
\mathbb{D}^2_{4}&=&
-128 \zf \ztw
+128 \ztw
-112 \zt
-76 \zt \zfr
+76 \zfr
-40 \zf
+5 \zsx
+(579/{2})\zs
-128, \nonumber\\
%%%
\mathbb{D}^3_{-5}&=&-256,\qquad
\mathbb{D}^3_{-4}=768,\qquad
\mathbb{D}^3_{-3}=512 \ztw,\qquad
\mathbb{D}^3_{-2}=
-768 \ztw
-384 \zt
-512, \nonumber\\
\mathbb{D}^3_{-1}&=&
-256 \ztw
+576 \zt
-416 \zfr
-768, \nonumber\\
\mathbb{D}^3_{0}&=&
448 \ztw \zt
+384 \zt
+80 \zfr
+192 \zf
-768, \nonumber\\
\mathbb{D}^3_{1}&=&
-336 \zt^2
-704 \ztw \zt
+192 \zt
-352 \zfr
+272 \zf
-1052 \zsx
-512, \nonumber\\
\mathbb{D}^3_{2}&=&
432 \zt^2
-832 \ztw \zt
+104 \zfr \zt
-256 \ztw
-336 \zfr
+1744 \ztw \zf
+1184 \zf\nonumber\\&&
+1616 \zsx
-1146 \zs, \nonumber\\
%%%%
\mathbb{D}^4_{-7}&=&-2560,\qquad
\mathbb{D}^4_{-6}=10240,\qquad
\mathbb{D}^4_{-5}=
7168 \ztw
-5120,  \nonumber\\
\mathbb{D}^4_{-4}&=&
-18432 \ztw
-5632 \zt
-10240, \nonumber\\
\mathbb{D}^4_{-3}&=&
14336 \zt
-13440 \zfr
-7680, \nonumber\\
\mathbb{D}^4_{-2}&=&
9216 \zt \ztw
+8192 \ztw
+1536 \zt
+14848 \zfr
+3200 \zf, \nonumber\\
%,
\mathbb{D}^4_{-1}&=&
-4416 \zt^2
-14080 \ztw \zt
-6144 \zt
+9216 \ztw
+3712 \zfr
-960 \zf
-7504 \zsx
+10240, \nonumber\\
\mathbb{D}^4_{0}&=&6464 \zt^2
-3328 \ztw \zt
-2400 \zfr \zt
-9728 \zt
+6144 \ztw
+1024 \zfr
-1664 \ztw \zf \nonumber\\&&
-4160 \zf
+({45008}/{3})\zsx
+2472 \zs
+20480, \nonumber\\
%%%%%
\mathbb{D}^5_{-9}&=&-28672,\qquad
\mathbb{D}^5_{-8}=143360,\qquad
\mathbb{D}^5_{-7}=
102400 \ztw
-143360,  \nonumber\\
\mathbb{D}^5_{-6}&=&
-368640 \ztw
-81920 \zt
-143360, \nonumber\\
\mathbb{D}^5_{-5}&=&
163840 \ztw
+292864 \zt
-289792 \zfr, \nonumber\\
\mathbb{D}^5_{-4}&=&
194560 \zt \ztw
+286720 \ztw
-110592 \zt
+622080 \zfr
+40960 \zf
+172032, \nonumber\\
\mathbb{D}^5_{-3}&=&
-87040 \zt^2
-473088 \ztw \zt
-241664 \zt
+184320 \ztw
+37888 \zfr
-46080 \zf \nonumber\\&&
+({137984}/{3})\zsx
+286720, \nonumber\\
\mathbb{D}^5_{-2}&=&
209408 \zt^2
+12288 \ztw \zt
-250368 \zfr \zt
-196608 \zt
-150528 \zfr
-24064 \ztw \zf \nonumber\\&&
-81920 \zf
+({767872}/{3})\zsx
+36736 \zs
+286720.
\end{eqnarray}

The coefficients in eq.~(\ref{DLgener}) are given by:
\begin{eqnarray}\label{App:2}
\mathfrak{C}^1_{0}&=&-16,\nonumber\\
\mathfrak{C}^1_{1}&=&16,\nonumber\\
\mathfrak{C}^1_{2}&=&16(1+ \ztw),\nonumber\\
\mathfrak{C}^1_{3}&=&16 (1-\zt),\nonumber\\
\mathfrak{C}^1_{4}&=&16 (1+\zfr),  \nonumber\\
\mathfrak{C}^1_{5}&=&16 (1-\zf),\nonumber\\
\mathfrak{C}^1_{6}&=&16 (1+ \zsx),\nonumber\\
\mathfrak{C}^1_{7}&=&16 (1- \zs),\nonumber\\
%\mathfrak{C}^1_{8}&=&16,\nonumber\\
%\mathfrak{C}^1_{9}&=&16,\nonumber\\
\mathfrak{C}^1_{8}&=&16 (1+ \ze),\nonumber\\
\mathfrak{C}^1_{9}&=&16 (1- \zn),\nonumber\\
\mathfrak{C}^2_{0}&=& - 64 \ztw,\nonumber\\
\mathfrak{C}^2_{1}&=&128 \ztw + 96 \zt,\nonumber\\
\mathfrak{C}^2_{2}&=&192 \ztw - 160 \zt - 8 \zfr,  \nonumber\\
\mathfrak{C}^2_{3}&=&
256 \ztw
-224 \zt
- 256 \ztw \zt
+ 152 \zfr
 + 360 \zf, \nonumber\\
\mathfrak{C}^2_{4}&=&
320 \ztw - 288 \zt + 144 \zt^2 + 216 \zfr - 144 \zf + (58/3) \zsx, \nonumber\\
\mathfrak{C}^2_{5}&=&
384 \ztw - 352 \zt + 280 \zfr - 280 \zt \zfr - 208 \zf - 384 \ztw \zf + 138 \zsx + 707 \zs,
 \nonumber\\
\mathfrak{C}^2_{6}&=&
448 \ztw - 416 \zt + 344 \zfr - 272 \zf + 202 \zsx - 134 \zs,
  \nonumber\\
\mathfrak{C}^2_{7}&=&
512 \ztw - 480 \zt + 408 \zfr - 336 \zf + 266 \zsx - 198 \zs,
  \nonumber\\
%%%
\mathfrak{C}^3_{0}&=&128 \zt + 256 \zfr,\nonumber\\
\mathfrak{C}^3_{1}&=&512 \zt + 1152 \ztw \zt + 672 \zfr - 960 \zf, \nonumber\\
\mathfrak{C}^3_{2}&=&
384 \zt - 2688 \ztw \zt - 1056 \zt^2 + 256 \zfr + 1504 \zf - (5000/3) \zsx, \nonumber\\
\mathfrak{C}^3_{3}&=&
-256 \zt - 3968 \ztw \zt + 1760 \zt^2 + 1248 \zfr - 1072 \zt \zfr + 4224 \zf + 6880 \ztw \zf \nonumber\\&& + (11696/3) \zsx - 6412 \zs,
 \nonumber\\
\mathfrak{C}^3_{4}&=&
- 1408 \zt - 5760 \ztw \zt + 2720 \zt^2 + 3648 \zf - 1904 \zt \zfr + 5920 \zf - 4800 \ztw \zf \nonumber\\&&  + (5080/3) \zsx + 2488 \zs,
\nonumber\\
\mathfrak{C}^3_{5}&=&
- 3072 \zt - 8064 \ztw \zt + 3936 \zt^2  + 7456 \zfr - 4048 \zt \zfr + 6592 \zf \nonumber\\&&  - 5952 \ztw \zf + 1664 \zsx + 9012 \zs,
 \nonumber\\
%%%%
\mathfrak{C}^4_{0}&=&2560 \ztw \zt + 384 \zt^2 - 128 \zf + (1888/3) \zsx,\nonumber\\
\mathfrak{C}^4_{1}&=&4608 \ztw \zt - 2944 \zt^2 + 7232 \zt \zfr - 10880 \zf - 20736 \ztw \zf - (70624/3) \zsx + 8848 \zs,  \nonumber\\
\mathfrak{C}^4_{2}&=&
3072 \ztw \zt - 7424 \zt^2 - 35136 \zt \zfr - 8704 \zf + 58880 \ztw \zf + (48512/3) \zsx - 704 \zs,  \nonumber\\
\mathfrak{C}^4_{3}&=&
- 8192 \ztw \zt - 5888 \zt^2 - 33408 \zt \zfr + 4352 \zf + 70400 \ztw \zf + (60224/3)  \zsx \nonumber\\&&
- 100912 \zs, \nonumber\\
%%%%%
\mathfrak{C}^5_{0}&=&
19456 \zt \zfr - 37888 \ztw \zf - 11520 \zs, \nonumber\\
\mathfrak{C}^5_{1}&=&
7168 \zt^2 + 19456 \zt \zfr - 100352 \ztw \zf + 189824 \zs.
\end{eqnarray}

For the ABA part of anomalous dimension the first three orders coincide with the result of full anomalous dimension. Other coefficients in eq.~(\ref{gammaDLABA}) are equal to:
\begin{eqnarray}\label{App:3}
%%%%
\overline{\mathbb{D}}^4_{-7}&=&-2560,\nonumber\\
\overline{\mathbb{D}}^4_{-6}&=&10240,\nonumber\\
\overline{\mathbb{D}}^4_{-5}&=&
7680 {\ztw}
-4608,\nonumber\\
\overline{\mathbb{D}}^4_{-4}&=&
-5888 {\zt}
-19456 {\ztw}
-10752, \nonumber\\
\overline{\mathbb{D}}^4_{-3}&=&
-17088 {\zfr}
+14848 {\zt}
-2560 {\ztw}
-9728, \nonumber\\
\overline{\mathbb{D}}^4_{-2}&=&
3840 {\zf}
+19584 {\zfr}
+12288 {\zt} {\ztw}
+4352 {\zt}
+8192 {\ztw}
-2048, \nonumber\\
\overline{\mathbb{D}}^4_{-1}&=&
-2240 {\zf}
+11968 {\zfr}
%-\frac{25520 {\zsx}}{3}
-(25520/3)\zsx
-5696 {\zt}^2
-18688 {\zt} {\ztw}
-6912 {\zt} \nonumber\\&&
+15360 {\ztw}
+11264, \nonumber\\
\overline{\mathbb{D}}^4_{0}&=&
1920 {\zf} {\ztw}
-1344 {\zf}
-15136 {\zfr} {\zt}
+11328 {\zfr}
+4264 {\zs}
+23504 {\zsx}
+8512 {\zt}^2 \nonumber\\&&
+18432 {\ztw}
-14080 {\zt} {\ztw}
-17408 {\zt}
+27648, \nonumber\\
%\end{eqnarray}
%%%%%%
%\begin{eqnarray}
\overline{\mathbb{D}}^5_{-9}&=&-28672,\nonumber\\
\overline{\mathbb{D}}^5_{-8}&=&143360,\nonumber\\
\overline{\mathbb{D}}^5_{-7}&=&
116736 {\ztw}
-129024,  \nonumber\\
\overline{\mathbb{D}}^5_{-6}&=&
-90112 {\zt}
-413696 {\ztw}
-174080, \nonumber\\
\overline{\mathbb{D}}^5_{-5}&=&
-418304 {\zfr}
+318464 {\zt}
+102400 {\ztw}
-55296, \nonumber\\
\overline{\mathbb{D}}^5_{-4}&=&
58880 {\zf}
+912128 {\zfr}
+289792 {\zt} {\ztw}
-40960 {\zt}
+393216 {\ztw}
+182272, \nonumber\\
\overline{\mathbb{D}}^5_{-3}&=&
-92160 {\zf}
+327168 {\zfr}
+181824 {\zsx}
-126976 {\zt}^2
-700416 {\zt} {\ztw} \nonumber\\&&
-350208 {\zt}
+458752 {\ztw}
+440320, \nonumber\\
\overline{\mathbb{D}}^5_{-2}&=&
-4096 {\zf} {\ztw}
-51200 {\zf}
-736768 {\zfr} {\zt}
-213504 {\zfr}
+64064 {\zs}
+241088 {\zsx} \nonumber\\&&
+306688 {\zt}^2
-269312 {\zt} {\ztw}
-466944 {\zt}
+221184 {\ztw}
+563200.
\end{eqnarray}

%\bibliographystyle{JHEP}

%\bibliography{WrappingDL45v4}

%\end{document}

%%%%%%%%%%%%%%%%%%%%%%%%%%%%%%%%%%%%%%%%%%%%%%%%%%%%%%%%%%%%
\end{document}